\documentstyle[floats,preprint,aps,epsf]{revtex}
\tightenlines

\newcommand{\del}{\partial}
\newcommand{\nn}{\nonumber}
\newcommand{\cov}{\bigtriangledown}

\newcommand{\beq}{\begin{equation}}
\newcommand{\eeq}{\end{equation}}
\newcommand{\beqa}{\begin{eqnarray}}
\newcommand{\eeqa}{\end{eqnarray}}
\newcommand{\sr}{\sqrt}
\newcommand{\fr}{\frac}
\newcommand{\mn}{\mu \nu}

\newcommand{\td}{\tilde{d}}
\newcommand{\vp}{\varphi}
\newcommand{\ve}{\varepsilon}
\newcommand{\D}{\Delta}

\begin{document}

\draft \preprint{ KEK-TH-832, hep-th/0209181}

\title{Classical Stability of Charged Black Branes and the
Gubser-Mitra Conjecture}
\author{ Takayuki Hirayama$^{a,}$\footnote{E-mail address:
hirayama@physics.utoronto.ca; Present address: Department of
Physics, University of Toronto, Toronto, Ontario, M5S1A7, Canada},
Gungwon Kang$^{a,}$\footnote{E-mail address: gwkang@kias.re.kr;
Present address: School of Physics, Korea Institute for Advanced
Study, 207-43 Cheongryangri-dong, Dongdaemun-gu, Seoul 130-012,
Korea}, and Youngone Lee$^{a,b,}$\footnote{E-mail address:
youngone@phya.yonsei.ac.kr}}
\address{$^a$Theory Group, KEK, Tsukuba, Ibaraki 305-0801, Japan \\
$^b$Dept. of Physics, Yonsei University, 134 Shinchon-dong, Seoul
120-749, Korea}

\maketitle
\begin{abstract}

We have investigated the classical stability of magnetically
charged black $p$-brane solutions for string theories that include
the case studied by Gregory and Laflamme. It turns out that the
stability behaves very differently depending on a coupling
parameter between dilaton and gauge fields. In the case of Gregory
and Laflamme, it has been known that the black brane instability
decreases monotonically as the charge of black branes increases
and finally disappears at the extremal point. For more general
cases we found that, when the coupling parameter is small, black
brane solutions become stable even before reaching to the extremal
point. On the other hand, when the coupling parameter is large,
black branes are always unstable and moreover the instability does
not continue to decrease, but starts to increase again as they
approach to the extremal point. However all extremal black branes
are shown to be stable even in this case. It has also been shown
that main features of the classical stability are in good
agreement with the local thermodynamic behavior of the
corresponding black hole system through the Gubser-Mitra
conjecture. Some implications of our results are also discussed.

\end{abstract}


\bigskip

\newpage

\section{Introduction}

Black holes are very important objects for studying and
understanding both general relativity and string theory. Brane
configurations in string theory have been studied from various
aspects, and play important roles in the context of gauge/gravity
duality \cite{Maldacena:1997re}. From the AdS/CFT correspondence
it has been also discussed that the IR limit of non supersymmetric
Yang Mills theories corresponds to black hole configurations in
AdS space which can be regarded as the near horizon limit of non
extremal D-brane configurations \cite{Witten:1998zw}. From the
study of Yang Mills theories we can reveal properties of black
holes, and it is discussed that the absence of tachyonic glueball
implies the stability of corresponding black hole configurations
\cite{Csaki:1998qr}. On the other hand, recently the issue about
the black string instability and its evolution has been of great
interest. It was widely believed that the evolution of the
instability in a black string results in bifurcations of the
horizon finally. However, Horowitz and Maeda have shown that the
horizon of a black string cannot pinch off in a finite affine time
\cite{Horowitz:2001cz}. Therefore an inhomogeneous black string
may be the final stable configuration. Recently some authors
discuss about the properties of inhomogeneous black strings and
their existence \cite{Gubser:2001ac,Horowitz:2002ym,wise}.

The Schwarzschild black hole is known to be stable under
linearized perturbations in general relativity \cite{VRW}.
However, Gregory and Laflamme discussed the stability of black
branes and found that a foliation of Schwarzschild black holes
(Sch.$\times$ S$^n$) is unstable if the compactification scale
(S$^n$ radius) is larger than the order of the Schwarzschild
radius -- the so-called Gregory-Laflamme instability
\cite{Gregory:vy}. One may, roughly speaking, expect that charge
prevents a black brane from being unstable because of some
repulsive forces among charges as the event horizon shrinks.
However they also showed that some charged black branes in ten
dimensional spacetime always have instability modes all the way
down to the extremal limit \cite{Gregory:1994bj}. So, it is
unlikely that the presence of charge is possible to remove the
instability in black branes they considered. On the other hand,
extremal black branes are shown to be stable
\cite{Gregory:1994tw}. The stability of black branes in anti-de
Sitter spacetime has also been investigated. In five-dimensional
anti-de Sitter spacetime, black strings that are foliations of
four-dimensional Schwarzschild black holes \cite{Gregory:2000gf}
or de Sitter-Schwarzschild black holes \cite{Hirayama:2001bi} are
recently shown to be unstable as well. However the black string
that is a foliation of four dimensional anti-de
Sitter-Schwarzschild black holes becomes stable as the horizon
radius is larger than the order of the AdS$_4$ radius
\cite{Hirayama:2001bi}. BTZ black strings in {\it four}
dimensional spacetime turn out to be stable always
\cite{Kang:2002hx}.

In string theory context, black branes that Gregory and Laflamme
considered \cite{Gregory:1994bj,Gregory:1994tw} are those having
magnetic charge with respect to Neveu-Schwarz gauge fields ({\it
i.e.}, the case of $\bar{\alpha} =-2$ in Eq.~(\ref{action2})
below). Thus, for example, black $p$-branes carrying charge with
respect to Ramond-Ramond gauge fields ({\it i.e.}, the case of
$\bar{\alpha} =0$) are not covered in their work. Properties of
D-branes and their non extremal extensions have many applications
and are important to study. Charged black brane solutions Gregory
and Laflamme considered might not be general enough to show other
possible interesting stability behavior. Therefore, it will be of
interest to see whether or not the stability behavior drastically
changes when more general types of charged black brane solutions
are considered. In fact we find different behaviors depending on
the coupling between dilaton and gauge fields. For instance
charged black branes  become stable even far from the extremal
point provided that this coupling is smaller than a certain
critical value. In this paper, we study the classical stability of
a wider class of charged black brane backgrounds which include
various black $p$-branes in string theory as special cases.

Another interesting feature of black hole systems is that they can
be regarded as thermodynamic systems. A black hole has entropy,
temperature, and other thermodynamic quantities. The area theorem
states that the total area of black holes cannot decrease, {\it
i.e.}, $\delta A \ge 0$, and it might be interpreted as the total
entropy cannot decrease $\delta S \ge 0$ in the black hole
thermodynamics. The behavior of instability explained above has
been commonly considered in the viewpoint of thermodynamics as a
reflection that a black string is entropically less-favorable
compared with multi black holes having the same mass and conserved
charges. However, the classical area theorem does not necessarily
require the transition of a black brane into a configuration
having larger horizon area. Thus, this global entropy argument is
too naive and cannot be applied to all cases. For example, one can
show that a black string in five-dimensional AdS space can be
stable even if its entropy is smaller than that of
five-dimensional AdS black holes \cite{Hirayama:2001bi}. As
pointed out by Reall \cite{Reall:2001ag}, the relevant one should
be the local thermodynamic behavior.

Recently, Gubser and Mitra gave such refinement of the entropy
argument and conjectured that a black brane with a non-compact
translational symmetry is classically stable if and only if it is
locally thermodynamically stable (GM conjecture) \cite{GM}. Reall
argued general correspondence between the classical instability
and the presence of a negative eigen mode in the Euclidean
Einstein-Hilbert action \cite{Reall:2001ag}. The negative eigen
mode gives an imaginary contribution to the path integral and then
it is closely related to the local thermodynamic instability. In
fact the local thermodynamic instability implies the existence of
the negative mode(s). However, there is no rigorous proof for the
converse: the existence of a negative mode indicates the
thermodynamic instability. For Schwarzschild
\cite{Prestidge:1999uq} and Reissner-Nordstr\"om \cite{GM} black
holes in AdS space and a Schwarzschild black brane enclosed in a
finite cavity \cite{Gregory:2001bd}, the equivalence between
non-existence of negative modes and local thermodynamic stability
are already examined by performing both classical and
thermodynamic stability analyses explicitly.

However, it is worthwhile to study classical stability of black
branes in the connection with black hole thermodynamics in a more
general context. Charged black brane solutions in type II
supergravity are particularly interesting. As pointed out by Reall
\cite{Reall:2001ag}, magnetically charged $p$-branes with $p \leq
4$ (D0, F1, D1, D2, D4) have sign changes in the specific heat as
the charge increases. Although in Ref.\,\cite{Reall:2001ag} the
essential feature of local thermodynamic stability for those black
branes has been analyzed and a sketch of the proof for the GM
conjecture is given in the context Reall considered, the actual
analysis for the classical stability has not been performed. So,
it has not been examined explicitly as yet how well classical and
thermodynamic stabilities actually agree with each other. This is
another main motivation of our study. By performing the classical
stability analysis, we give an explicit check for the validity of
the GM conjecture. As shall be shown below, the agreement between
our numerical results for classical perturbations and the
thermodynamic stability through the GM conjecture is very good. We
also show some features in the classical stability behavior that
are not easily predicted by the thermodynamic behavior only.

This paper is organized as follows. In section \ref{thermo}, we
summarize the thermodynamic stability behavior of black branes we
are considering. Section \ref{linear} is devoted to the numerical
analysis for the classical stability. We give conclusions and
discussions in the last section.

\section{Local thermodynamic stability of Black $p$-branes}
\label{thermo}

In general, it is not easy to check whether or not black string or
brane background spacetimes are stable under classical linearized
perturbations. Such classical analysis involves numerical
calculations for most cases. As stated above briefly, however,
there has been suggested recently a simple method from which one
can predict about the basic behavior of the classical stability.
This is the so-called Gubser-Mitra (GM) conjecture~\cite{GM}. In
this section, before performing the classical analysis in detail,
we first consider the local thermodynamic behavior of a finite
segment of a black $p$-brane background spacetime.

We consider black $p$-brane solutions of the following action
\beq
{\rm I} = \int d^Dx\sr{-\bar{g}} \left[ e^{-\bar{\beta}
\bar{\phi}} \left( \bar{R} -\bar{\gamma} (\partial \bar{\phi})^2
\right) -\fr{1}{2n!} e^{\bar{\alpha}\bar{\phi}}F_n^2 \right] .
\label{action2}
\eeq
Here $F_n$ denotes a $n$-form field strength.
$\bar{\beta}$, $\bar{\gamma}$, and $\bar{\alpha}$ are assumed to
be arbitrary constants with $\hat{\gamma} = \bar{\gamma}
+(D-1)\bar{\beta}^2/(D-2) > 0$. However, by taking the conformal
transformation\footnote{Note that the classical stability of black
branes does not change under a conformal transformation as long as
the conformal factor is regular. Moreover, black hole
thermodynamic quantities such as temperature \cite{Jacobson:pf},
entropy, and specific heat are invariant under stationary
conformal transformations.} $\bar{g}_{MN}
=e^{2\bar{\beta}\bar{\phi}/(D-2)}\hat{g}_{MN}$ and then rescaling
the dilaton field $\bar{\phi}=\phi/\sr{2\hat{\gamma}}$, this
action can be written as
\beq
{\rm I} = \int d^Dx\sr{-\hat{g}}
\left[ \hat{R} -\fr{1}{2} (\partial \phi)^2
-\fr{1}{2n!}e^{a\phi}F_n^2 \right] , \label{action1} \eeq where
\beq a=\fr{\bar{\alpha} +(D-2n)\bar{\beta}
/(D-2)}{\sr{2\hat{\gamma}}} .
\label{conftrans}
\eeq

Note that the sign of the parameter $a$ is not relevant here since
the action is invariant under $a \rightarrow -a$ and $\phi
\rightarrow -\phi$. In the following analysis, we will consider
only magnetically charged black brane solutions of this action,
not electrically charged ones. It is because electrically charged
solutions can be obtained from magnetically charged ones by
dualizing the $n$-form field $F$. Moreover, the magnetically
charged case is technically easier to treat in the perturbation
analysis.\footnote{ Gauge perturbations can be set to be zero for
the sphere directions, and are decoupled from metric and scalar
perturbations for other directions.} The metric of magnetically
charged black $p$-brane solutions is given by
\cite{Horowitz:cd,Duff:1996hp}
\beqa
d\hat{s}^2 &=& -\left(
1+\fr{k}{r^{\tilde{d}}} \sinh^2 \mu \right)^{-\fr{4\td} {\triangle
(D-2)}} U dt^2 +\left( 1+\fr{k}{r^{\tilde{d}}} \sinh^2 \mu
\right)^{\fr{4d}{\triangle (D-2)}} \left( \fr{dr^2}{U} +r^2
d\Omega^2_{n} \right) \nn  \\
& &  +\left( 1+\fr{k}{r^{\tilde{d}}} \sinh^2 \mu
\right)^{-\fr{4\td}{\triangle (D-2)}} \delta_{ij}dz^idz^j , \nonumber \\
e^{-\fr{\triangle}{2a}\phi} &=& 1+\fr{k}{r^{\tilde{d}}} \sinh^2
\mu , \qquad\qquad U= 1-\fr{k}{r^{\td}}, \qquad\qquad \triangle
=a^2 +\fr{2\td d}{D-2} .
\label{metric}
\eeqa
The $n$-form field
strength is proportional to the volume form on $S^n$. Here $\td
=n-1$, $d=p+1$, $D=\td +d +2 = 2+n+p$, and the coordinates are $\{
x^M\} =\{ x^{\mu}, z^i\}=\{ t,r,x^m,z^i \}$ with $m=1,\cdots, n$
and $i=1,\cdots ,p$. Thus, the coordinates $\{z^i\}$ describe the
$p$-dimensional spatial worldvolume, and $\{x^m\}$ the $n$-sphere.
The spatial worldvolume is not compactified in our consideration.

Note that the metric components for the $p$-dimensional spatial
worldvolume directions above have an overall multiplication factor
depending on the $r$-coordinate. This multiplication factor makes
the spatial worldvolume non-flat, and turns out to give some
complications in the perturbation analysis. As pointed out by
Reall \cite{Reall:2001ag}, such complications can be easily
avoided by performing an appropriate conformal transformation
again such as $\hat{g}_{MN}=e^{2(n-1)\phi/(D-2)a}g_{MN} $. Then
the action becomes
\beq
{\rm I} = \int d^Dx\sr{-g} \left[
e^{-\beta\phi} \left( R -\gamma (\partial \phi)^2 \right)
-\fr{1}{2n!} e^{\alpha\phi}F_n^2 \right] ,
\label{action3}
\eeq
where the constants are now functions of a single parameter $a$ as
follows:
\beq
\beta = -\fr{n-1}{a}, \qquad \gamma = \fr{1}{2}
-\fr{(D-1)(n-1)^2} {(D-2)a^2}, \qquad \alpha =
a+\fr{(D-2n)(n-1)}{(D-2)a} .
\eeq
And the metric for the black
$p$-brane background can be written as
\beq
ds^2 = -Udt^2
+V^{-1}dr^2 +R^2d\Omega^2_n +\delta_{ij}dz^idz^j ,
\label{conmetric}
\eeq
where
\beq V^{-1} = \fr{\left(
1+\fr{k}{r^{\tilde{d}}}\sinh^2 \mu
\right)^{\fr{4}{\triangle}}}{1-k/r^{\td}}, \qquad R^2 = \left(
1+\fr{k}{r^{\tilde{d}}} \sinh^2 \mu \right)^{\fr{4} {\triangle}}
r^2  .
\eeq
The mass $M$ and magnetic charge $Q$ per unit
$p$-volume are given by \cite{Duff:1996hp}
\beq
M = k (\td +1
+\fr{4\td}{\triangle}\sinh^2\mu ) \qquad {\rm and} \qquad Q =
\fr{\td k}{\sr{\triangle}} \sinh 2\mu ,
\label{MC}
\eeq
respectively.

Note that the case of $\bar{\beta}=2$, $\bar{\gamma}=-4$,
$\bar{\alpha}=-2$, and $D=10$ in Eq.~(\ref{action2}) corresponds
to the action for which the classical stability of magnetically
charged black $p$-brane solutions has been studied by Gregory and
Laflamme (GL)~\cite{Gregory:1994bj,Gregory:1994tw}. Thus, as
pointed out by Reall~\cite{Reall:2001ag}, NS$5$-brane ({\it i.e.},
$n=3$) is the only black brane of the type II supergravity covered
by their work. This case corresponds to $a=(1-n)/2$ in the
conformally transformed action in Eq.~(\ref{action3}). F-string by
dualizing the Neveu-Schwarz gauge field can also be described by
the action in Eq.~(\ref{action2}) with a different coupling to the
dilaton, {\it i.e.}, $\bar{\alpha} =2$ and so $a=(9-n)/2$. Hence
F-string ({\it i.e.}, $n=7$) can be covered. When $\bar{\beta}=2$,
$\bar{\gamma}=-4$, $\bar{\alpha}=0$, and $D=10$, it gives
$a=(5-n)/2$. Solutions in Eq.~(\ref{conmetric}) for this case
cover D$p$-branes with $p=0$, $1$, $2$, $4$, $5$, $6$ of the type
II supergravity carrying magnetic charge. Therefore, our study
contains a broad class of black brane solutions, including most of
black branes in the type II supergravity.

Now let us consider a black $p$-brane with unit worldvolume. Being
regarded as a thermal system, it has entropy and temperature given
by
\beq
S \sim \left( \cosh \mu \right)^{\fr{4}{\triangle}}
r_{\scriptscriptstyle{H}}^{n}  \quad {\rm and} \quad
T=\fr{\tilde{d}}{4 \pi r_{\scriptscriptstyle{H}}} \left( \cosh\mu
\right)^{-\fr{4}{\D}} ,
\label{ST}
\eeq
respectively. Here
$r_{\scriptscriptstyle{H}} = k^{1/\td}$ is the horizon radius. The
extremal limit of the magnetically charged black brane solution is
$k \rightarrow 0$ and $\mu \rightarrow \infty $ with keeping mass
and charge in Eq.\,(\ref{MC}) finite ({\it i.e.}, $k e^{2\mu}$
fixed). Note that in the extremal limit both temperature and
entropy go to zero except for the case of $\triangle =2\td$,
\footnote{For the case of $\triangle =2\td$, interestingly the
temperature does not vanish although the entropy still goes to
zero in the extremal limit.} and the solution becomes the B.P.S.
state which is known to be stable even at the quantum level. In
order for this system to be stable thermodynamically, the entropy
functional given above should be a local maximum in thermodynamic
phase space. Defining the Hessian of the system as
\beq
H = \left(
\matrix{  \fr{\del ^2 S}{\del M^2} & \fr{\del ^2 S}{\del M \del Q}
\cr \fr{\del ^2 S}{\del M \del Q} & \fr{\del ^2 S}{\del
Q^2}}\right) ,
\eeq
the maximum entropy is guaranteed provided
that
\beq
\left( \fr{\del ^2 S}{\del M^2} \right)_Q < 0  \qquad
{\rm and} \qquad  \det \left( H \right) > 0 ,
\eeq
which become
equivalent to~\cite{Landau}
\beq
\label{StableCondition}
C_Q
=\left( \fr{\del M}{\del T}\right)_Q >0 \qquad  {\rm and} \qquad
\left(\fr{\del \Phi_{\scriptscriptstyle{H}}}{\del Q} \right)_T >0.
\eeq
Here $\Phi_{\scriptscriptstyle{H}}$ denotes the magnetic
potential energy at the horizon. Using Eqs.~(\ref{MC},\ref{ST}),
we obtain explicitly
\beqa
\label{HeatCapacity}
\mbox{} & & C_Q =
-4 \pi r_{\scriptscriptstyle{H}}^{\tilde{d}+1} \left( \cosh \mu
\right)^{\fr{4}{\Delta}}
           \fr{2\tilde{d}+(\Delta+\tilde{d}(\Delta -2))\cosh 2\mu}{2
\tilde{d} +(\Delta-2\tilde{d})\cosh 2\mu} , \\
\label{PotentialCapacity}
& & \left(\fr{\del \Phi_{\scriptscriptstyle{H}}}{\del Q}\right)_T
= \fr{ r_{\scriptscriptstyle{H}}^{- \tilde{d}}
\Delta \cosh 2\mu}{2 \tilde{d} +(\Delta-2\tilde{d})\cosh 2\mu} .
\eeqa

Since the term $\Delta+\tilde{d}(\Delta -2)$ is positive definite,
one can find that $C_Q$ and $(\del
\Phi_{\scriptscriptstyle{H}}/\del Q)_T$ always have opposite signs
in the above. Consequently, two conditions in
Eq.~(\ref{StableCondition}) cannot be satisfied simultaneously. In
other words, the black brane system in the consideration cannot be
thermodynamically stable for processes in which both mass $M$ and
charge $Q$ vary. Therefore, according to the GM conjecture, this
may indicate that the black brane background is always unstable
under classical perturbations. As mentioned by
Reall~\cite{Reall:2001ag}, however, the charge of the black brane
cannot fluctuate since there is no matter field carrying charge in
our theory. Thus, only thermodynamic processes through which the
charge does not change are relevant. Accordingly, the local
thermodynamic stability of a black brane is determined solely by
the sign of the specific heat $C_Q$ above.

The specific heat in Eq.~(\ref{HeatCapacity}) is always negative
if $\Delta -2\td \geq 0$, equivalently, $|a| \geq a_{\rm cr}$
where the critical value of $a$ depends on the form $n$ of the
gauge field, or the dimension $p$ of the spatial worldvolume as
\beq
a_{\rm cr} = \fr{n-1}{\sr{(D-2)/2}} =
\fr{D-3-p}{\sr{(D-2)/2}} .
\label{criticala}
\eeq
For $|a| <a_{\rm
cr}$, however, the specific heat is negative if $0 \leq \mu <
\mu_{\rm cr} $, but positive if $\mu_{\rm cr} < \mu < \infty$. At
$\mu=\mu_{\rm cr}$, it diverges, {\it i.e.}, $C_{Q} \rightarrow
\mp \infty$ as $\mu \rightarrow \mu_{\rm cr} \mp 0$. Here the
critical value of the parameter $\mu$ is
\beq
\sinh^2\mu_{\rm cr}
= \fr{2(D-3-p)(p+1)+(D-2)a^2}{2\left[ 2(D-3-p)^2
-(D-2)a^2\right]}.
\label{criticalmu}
\eeq
Based on the GM
conjecture, therefore, it can be expected under classical
linearized perturbations that a black $p$-brane background is
always unstable provided $|a| \geq a_{\rm cr}$, {\it independent
of its charge}. For $|a| < a_{\rm cr}$, however, it becomes stable
for $\mu > \mu_{\rm cr}$. For instance, black $p$-branes
considered by Gregory and
Laflamme~\cite{Gregory:1994bj,Gregory:1994tw} have $a=(1-n)/2$ as
explained above. This value is exactly the same as the critical
value for $D=10$ in Eq.~(\ref{criticala}), {\it i.e.},
$a^2=a^2_{\rm cr}$, implying negative specific heat always. Thus
it is expected that all black branes considered by them, including
NS$5$-brane, are unstable, and this behavior is shown by them. For
F-string, we have $a=1 < a_{\rm cr}=3$. For D$p$-brane solutions
of the type II string theory, we have $a=(p-3)/2$ and $a_{\rm
cr}=(7-p)/2$. Therefore, one can easily see that the specific heat
changes sign for magnetically charged D$0$, F$1$, D$1$, D$2$, D$4$
branes, but not for D$5$, D$6$ branes as mentioned by
Reall~\cite{Reall:2001ag}. The case of D3-branes will not be
considered in this paper since the action for the self dual gauge
field is not known. Moreover, even if we simply use the action in
Eq.~(\ref{action2}) and impose the self duality condition, we
encounter some problems for analyzing linearized perturbation
equations. Finally we would like to point out that, except for
cases of $|a|=a_{\rm cr}$, the specific heat goes to zero in the
extremal limit, satisfying a formulation of the ordinary
thermodynamic third law.

\section{Classical perturbation analysis}
\label{linear}

So far, we have analyzed the local thermodynamic stability of
black $p$-brane systems that gives some hints at the stability of
such background spacetimes through the GM conjecture. Now let us
perform the classical stability analysis explicitly. We consider
small metric perturbations about black $p$-brane background
spacetimes and see whether or not there exists any mode that is
regular spatially, but grows exponentially in time.

Since the background black $p$-brane spacetime in
Eq.~(\ref{conmetric}) is independent of the time coordinate $t$
and spatial worldvolume coordinates $z^i$, one can assume that
\beqa
\mbox{} & & \delta g_{MN}=h_{MN}(x^{\mu}, z^i) =e^{\Omega
t+im_iz^i}H_{MN}(r,x^m),  \nn \\
& & \delta \phi =e^{\Omega t+im_iz^i}f(r,x^m), \qquad \delta F
=e^{\Omega t+im_iz^i}\delta F(r,x^m)
\eeqa
for unstable mode
solutions. Here $\Omega > 0$ and the Kaluza-Klein (KK) mass $m_i$
is a continuous real number. As explained in
Refs.~\cite{Kang:2002hx,Reall:2001ag}, one can further use the
diffeomorphism symmetry so that the scalar and vector parts of
metric fluctuations $h_{MN}$ are set to be zero for {\it massive}
modes in the viewpoint of the Kaluza-Klein
reduction~\cite{footnote2}
\beq
H_{\mu i}=H_{ij}=0.
\eeq
Since in
general non-$s$-wave fluctuations are expected to be more stable
than $s$-wave ones, we consider only spherically symmetric
perturbations.\footnote{We will give an argument below that the
consideration of $s$-wave perturbations only might be enough to
show stability.} Thus, as shown in Refs.~\cite{VRW}, the following
gauge choice can be made
\beq
H_{tm}=H_{rm}=0, \qquad\qquad  H^m_n
=K(r)\delta^m_n
\eeq
for such perturbations. Other components
$H_{tt}, H_{tr}, H_{rr}$ and the dilaton and field strength
perturbations are also independent of the angular coordinates
$x^m$. That is, they are functions of the radial coordinate $r$
only.

For these forms of perturbations above, one can easily show that
linearized equations for gauge fluctuations $\delta F$ are
decoupled, {\it i.e.}, $\cov_N (e^{\alpha\phi}\delta F^{NP_1\cdots
P_{n-1}}) =0$. Moreover, as in Ref.~\cite{Gregory:1994bj}, it can
be shown that for background spacetimes in Eq.~(\ref{conmetric})
there exists no unstable mode solution for the spherically
symmetric fluctuations. One can first show that the fluctuations
proportional to the volume form on $S^n$ should vanish if $\Omega
\not= 0$. Thus only remaining components $\delta F_{ti_1\cdots
i_{n-1}}$, $\delta F_{ri_1\cdots i_{n-1}}$, $\delta F_{tri_1\cdots
i_{n-2}}$, and $\delta F_{i_1\cdots i_n}$ of such fluctuations
could be non-zero. Here $i_1, \cdots, i_{n-1}$ are the spatial
worldvolume coordinates ${\bf z}$. By using the Bianchi identity
$\cov_{[N }\delta F_{P_1\cdots P_n]}=0$ for $(NP_1\cdots
P_n)=(tri_1\cdots i_{n-1})$ and the perturbation equations for
$(P_1\cdots P_{n-1})=(ri_1\cdots i_{n-2}), (ii_1\cdots i_{n-2})$
above, one obtain coupled equations for ${\bf f}_t \equiv \sum_i
m_i\delta F_{tii_1\cdots i_{n-2}}$, ${\bf f}_r \equiv \sum_i
m_i\delta F_{rii_1\cdots i_{n-2}}$, and $\delta F_{tri_1\cdots
i_{n-2}}$ that can be easily decoupled as
\beq
V {\bf f}''_t
+\left( \cdots \right) {\bf f}'_t
-\left(m^2+\fr{\Omega^2}{U}\right) {\bf f}_t =0 ,
\eeq
where $m^2
= \sum_i m_i^2$. Since both $U$ and $V$ are positive outside the
horizon, one can see that the only regular solution of this
equation is $ {\bf f}_t=0$, giving $ {\bf f}_r =\delta
F_{tri_1\cdots i_{n-2}}=0$ correspondingly. Similarly, one also
finds that there are no non-vanishing regular solutions for other
components of the fluctuations. Therefore we set $\delta F=0$ in
the following analysis.

As shown in Ref.~\cite{Reall:2001ag}, now the dilaton perturbation
equation is given by
\beqa
\mbox{} & & \cov^2 f -2\beta g^{\mn}\cov_{\mu}\phi \cov_{\nu}
f -H^{\mn}\cov_{\mu}\cov_{\nu}\phi +\beta H^{\mn}\cov_{\mu}
\phi \cov_{\nu}\phi -\cov_{\mu}\phi \cov_{\nu} \left( H^{\mn}
-\fr{1}{2}H^{\rho}_{\rho} g^{\mn} \right)   \nn  \\
& & \qquad  +\fr{a}{2(n-1)!} e^{(\alpha +\beta )\phi}
\left[ H^{\mn}F_{\mu\rho_1\cdots\rho_{n-1}}
{F_{\nu}}^{\rho_1\cdots\rho_{n-1}} -\fr{\alpha +\beta}{n}F^2
f \right] = m^2 f .
\label{delphi2}
\eeqa
And linearized equations for metric perturbations $H_{\mn}$ are given by
\beqa
\mbox{} & & \cov^2 H_{\mn} -2\cov_{(\mu}\cov^{\rho}
H_{\nu )\rho} +\cov_{\mu}\cov_{\nu} H^{\rho}_{\rho}
-2R_{\rho (\mu}H^{\rho}_{\nu )} +2R_{\mu\rho\nu\sigma}
H^{\rho\sigma}  \nn  \\
& & \qquad  +\beta \left( 2\cov_{(\mu}H^{\rho}_{\nu )}
-\cov^{\rho}H_{\mn} \right) \cov_{\rho}\phi -2\beta
\cov_{\mu}\cov_{\nu} f +4(\gamma +\beta^2)\cov_{(\mu}
\phi \cov_{\nu )} f     \nn  \\
& & \qquad  -\fr{1}{(n-1)!} e^{(\alpha +\beta )\phi}
\left[ (n-1)H^{\rho\sigma}
F_{\mu\rho\lambda_1\cdots\lambda_{n-2}}{F_{\nu\sigma}}
^{\lambda_1\cdots\lambda_{n-2}} -(\alpha +\beta ) f
F_{\mu\lambda_1\cdots\lambda_{n-1}}{F_{\nu}}
^{\lambda_1\cdots\lambda_{n-1}} \right]  \nn  \\
& &   \qquad = m^2 H_{\mn} .
\label{delgmn}
\eeqa
The $\{\mu i \}$ and $\{ ij \}$ components are
\beqa
\label{delgmi}
\nabla_{\nu}H^{\nu}_{\mu} -\beta H^{\nu}_{\mu}
\nabla_{\nu}\phi -2(\gamma +\beta^2) f\nabla_{\mu}\phi =0, \\
H^{\rho}_{\rho} -2\beta f =0 ,
\label{delgij}
\eeqa
respectively.
Here the covariant derivative and curvature tensors in the above
equations are constructed by the metric $g_{\mn}$ only. In other
words, perturbation equations on $D$-dimensional black $p$-brane
backgrounds become coupled second order differential equations for
dilaton and gravitational fields with mass $m$ on
$(D-p)$-dimensional black hole backgrounds, which is expected from
the viewpoint of Kaluza-Klein dimensional reductions.

As pointed out in Ref.~\cite{Reall:2001ag} already, notice also
that the dilaton perturbation equation Eq.~(\ref{delphi2}) is
actually not independent, but can be obtained from the trace of
Eq.~(\ref{delgmn}) and Eqs.~(\ref{delgmi},\ref{delgij}). In
Eq.~(\ref{delgij}), one finds that the dilaton fluctuation $f$ can
be computed once the metric fluctuations $H_{\mn}$ are obtained.
Thus, we need to consider only the metric perturbation variables,
which are $H_{tt}$, $H_{tr}$, $H_{rr}$, and $K$. For the stability
analysis, however, it suffices to consider the behavior of the
so-called threshold modes, {\it i.e.}, KK massive modes having
$\Omega =0$. This is because infinitely heavy massive modes cannot
be expected to give instability. There should be a maximum value
$m^*$ for the instability to exist. For unstable modes in the case
of black brane systems, if they exist, it is known that the
corresponding ``imaginary'' frequency $\Omega$ usually starts to
increase as the KK mass increases, reaches to some maximum value,
but finally decreases to zero at the so-called threshold mass
$m^*$. Thus, as long as the set of equations in
Eqs.~(\ref{delgmn}-\ref{delgij}) allows a threshold mode solution
with non-vanishing mass, the instability exists. When $\Omega =0$,
there is an additional merit that a further gauge choice is
possible to set $H_{tr}=0$ as in Ref.~\cite{Reall:2001ag}.
Therefore, one finally has
\beq
\left( H^{\mu}_{\nu} \right) =
{\rm diag}\left( \varphi (r), \psi (r), \chi (r), \cdots , \chi
(r) \right) .
\eeq
That is, the perturbed metric for threshold
modes can be expressed in the following form
\beq
ds^2 = -U \left(
1+\vp e^{im_iz^i} \right) dt^2 +V^{-1}\left( 1+\psi e^{im_iz^i}
\right) dr^2 +R^2\left( 1+\chi e^{im_iz^i} \right) d\Omega^2_n
+d{\bf z}^{2}.
\label{pertmetric}
\eeq

Now Eqs.~(\ref{delgmn}-\ref{delgij}) reduce to
\beqa
\label{varphi}
\mbox{} & & \vp'' +\left( \fr{U'}{2U} +\fr{V'}{2V} +n\fr{R'}{R}
-\beta \phi' \right) \vp' +\left[ \fr{U''}{U} -\fr{U'}{U}\left(
\fr{U'}{2U} -\fr{V'}{2V} -n\fr{R'}{R} +\beta \phi' \right)
-\fr{m^2}{V} \right] \vp  \nn  \\
&& \qquad -\fr{U'}{U} \psi' -\left[ \fr{U''}{U} -\fr{U'}{U}
\left( \fr{U'}{2U} -\fr{V'}{2V} -n\fr{R'}{R} +\beta \phi' \right)
\right] \psi =0 , \\
\label{psi}
&& \psi'' +\left( \fr{U'}{2U} +\fr{V'}{2V} +n\fr{R'}{R}
-\fr{2\gamma +3\beta^2}{\beta} \phi' \right) \psi' +\fr{m^2}{V}
\psi -\left( \fr{U'}{U} +2\fr{\gamma +\beta^2}{\beta} \phi'
\right) \vp'   \nn  \\
&& \qquad -2n\left( \fr{R'}{R}+\fr{\gamma +\beta^2}{\beta} \phi'
\right) \chi' =0 ,  \\
\label{chi}
&& \psi' +\left( \fr{U'}{2U} +n\fr{R'}{R} -\fr{\gamma +2\beta^2}
{\beta} \phi' \right) \psi -\left( \fr{U'}{2U} +\fr{\gamma +\beta^2}
{\beta} \phi' \right) \vp -n\left( \fr{R'}{R} +\fr{\gamma +\beta^2}
{\beta} \phi' \right) \chi =0 .
\eeqa
Using Eq.~(\ref{chi}), one can easily see that $\chi$ can be
evaluated from $\vp$ and $\psi$, and that Eqs.~(\ref{varphi},\ref{psi})
become two second order coupled equations for $\vp$ and $\psi$ only
as follows
\beqa
\label{varphif}
\mbox{} && r\left( r^{\td} -k \right) \vp'' +\left[ (\td +1)r^{\td}
-k \right] \vp' -m^2r^{\td +1-4\td /\triangle} \left( r^{\td}
+k\sinh^2\mu \right)^{4/\triangle} \vp -\td k \psi' =0 , \\
\label{psif}
&& r^2\left( r^{\td} -k \right)^2 \left( r^{\td} +k\sinh^2\mu \right)
\psi'' +r \left( r^{\td} -k \right)^2 \Bigg[ 2\td \left( r^{\td}
-\fr{2}{\triangle} k\sinh^2\mu \right)   \nn  \\
&& \qquad -(\td -3)\left( r^{\td} +k\sinh^2\mu\right) \Bigg] \psi'
-\Bigg\{ m^2r^{\td +2-4\td /\triangle}\left( r^{\td}-k \right)
\left( r^{\td} +k\sinh^2\mu \right)^{1+4/\triangle}  \nn  \\
&& \qquad  +\td k \left[ W +\fr{2}{\triangle} \left( 2{\td}^2
+(\td +3)(a^2 -\fr{2{\td}^2}{D-2}) \right) \sinh^2\mu \left(
r^{\td} -k \right)^2 \right] \Bigg\} \psi +\td k W \vp =0 .
\eeqa
Here the functions $U$, $V$, and $R$ are substituted explicitly,
and
\beq
W= \td \left( r^{\td}-k \right)\left( r^{\td}
+k\sinh^2\mu \right) -\fr{2}{\triangle}\left( a^2
-\fr{2{\td}^2}{D-2}\right) \sinh^2\mu \left( r^{\td} -k \right)^2
+\td k\cosh^2\mu \,r^{\td} .
\eeq
The question about whether the
black $p$-branes are stable or unstable under linearized
perturbations now becomes whether or not there exists some
Kaluza-Klein mass parameter $m$ for which the above coupled
equations allow any regular solution outside the event horizon.

Before going further, several points should be mentioned here. One
may wonder if metric fluctuations alone with frozen dilaton
perturbation could produce instability. When $f=0$,
Eq.~(\ref{delgij}) gives $-n\chi = \vp +\psi$, and so from
Eq.~(\ref{chi}) one has an additional first order equation for
$\vp$ and $\psi$. This raises the question about whether or not
such equation is consistent with Eqs.~(\ref{varphif},\ref{psif}).
It turns out that the inconsistency is proportional to
$\sinh^2\mu$. We find therefore that metric fluctuations alone
without dilaton perturbation cannot produce linearized instability
when black branes are charged, {\it i.e.}, $\mu \not= 0$. It is
possible only for uncharged black branes, {\it i.e.}, $\mu =0$.
This property can also be expected by applying the same argument
as in the work of Gregory and Laflamme~\cite{Gregory:1994bj}. It
is shown in Ref. \cite{Gregory:1994bj} that for small charge cases
$f=f_0 +\vartheta (\sinh^2\mu)$ where $f_0$ is independent of
$\mu$. Now it can be easily shown in Eq.~(\ref{delphi2}) that
$f_0=0$.

Secondly, Eqs.~(\ref{varphif},\ref{psif}) are invariant under
rescalings of $m \rightarrow \alpha m$, $r_{\scriptscriptstyle{H}}
\rightarrow r_{\scriptscriptstyle{H}}/\alpha$, $\mu \rightarrow
\mu$, and $r \rightarrow r/\alpha$ for an arbitrary constant
$\alpha$. This scaling symmetry can be easily seen by defining
dimensionless variables such as $\rho
=r/r_{\scriptscriptstyle{H}}$ and $\bar{m}=m
r_{\scriptscriptstyle{H}}$, or by observing that such rescalings
with $t \rightarrow t/\alpha$ and $z^i \rightarrow z^i/\alpha$
simply result in an overall constant rescale of the metric under
which field equations are not affected. This symmetry implies that
the threshold mass must be inversely proportional to the horizon
radius, {\it i.e.}, $m^* \sim 1/r_{\scriptscriptstyle{H}}$.
Accordingly, it suffices to study instability modes just for a
single value of $r_{\scriptscriptstyle{H}}$.

Since perturbation equations above are coupled second order and
linear in $\psi$ and $\vp$, there are four linearly independent
mode solutions in general. The asymptotic solutions of
Eqs.~(\ref{varphif},\ref{psif}) at the spatial infinity ({\it
i.e.}, $r \sim \infty$) are given by
\beqa
\vp (r) & \simeq &
e^{\pm mr}u_{\pm}(r) \simeq e^{\pm mr} \left[ r^{-\fr{\td +1}{2}}
\mp \fr{(\td -1)(\td +1)}{8m} r^{-\fr{\td
+3}{2}} +\cdots \right] , \\
\psi (r) & \simeq & e^{\pm mr}v_{\pm}(r) \simeq e^{\pm mr} \left[
r^{-\fr{\td +3}{2}} \mp \fr{(\td +1)(\td +5)}{8m} r^{-\fr{\td
+3}{2}} +\cdots \right]
\label{inftysol}
\eeqa
up to overall
arbitrary constants. By finding out asymptotic solutions in the
vicinity of the event horizon as well, {\it i.e.}, $\rho =
1+\varepsilon$, one can have four sets of mode solutions whose
asymptotic behaviors are given by
\beqa
\label{solI}
\psi_{\scriptscriptstyle{I}} & \sim & \left\{\begin{array}{l}
                    \ve +\vartheta (\ve^2)  \\
A_{\scriptscriptstyle{I}} e^{-\bar{m}\rho} v_-(\rho)
+B_{\scriptscriptstyle{I}}e^{\bar{m}\rho} v_+(\rho) ,
                    \end{array}
             \right.    \quad\quad
\psi_{\scriptscriptstyle{II}} \sim \left\{\begin{array}{l}
                    1 +\vartheta (\ve^2)  \\
A_{\scriptscriptstyle{II}} e^{-\bar{m}\rho} v_-(\rho)
+B_{\scriptscriptstyle{II}}e^{\bar{m}\rho}v_+(\rho),
                   \end{array}
             \right.  \\
\label{solIII}
\psi_{\scriptscriptstyle{III}} & \sim & \left\{\begin{array}{l}
                    \ve\ln\ve +\vartheta (\ve^2\ln\ve) \\
A_{\scriptscriptstyle{III}} e^{-\bar{m}\rho} v_-(\rho)
+B_{\scriptscriptstyle{III}}e^{\bar{m}\rho}v_+(\rho),
                    \end{array}
             \right.  \,\,\,
\label{solIV} \psi_{\scriptscriptstyle{IV}} \sim
\left\{\begin{array}{ll} \ve^{-1}
-\fr{\bar{m}^2}{\td}(\cosh\mu)^{8/\triangle}\ln\ve
+\vartheta (\ve^2)\\
A_{\scriptscriptstyle{IV}} e^{-\bar{m}\rho} v_-(\rho)
+B_{\scriptscriptstyle{IV}}e^{\bar{m}\rho}v_+(\rho) ,
                    \end{array}
             \right.
\eeqa
and correspondingly
\beqa
\label{solIvp}
\vp_{\scriptscriptstyle{I}} & \sim & \left\{\begin{array}{l}
                    \ve +\vartheta (\ve^2)  \\
\bar{A}_{\scriptscriptstyle{I}} e^{-\bar{m}\rho} u_-(\rho)
+\bar{B}_{\scriptscriptstyle{I}}e^{\bar{m}\rho} u_+(\rho) ,
                    \end{array}
             \right.  \quad\,\,\,\,\,\,\,
\vp_{\scriptscriptstyle{II}} \sim \left\{\begin{array}{l}
1 +\fr{\bar{m}^2}{\td}(\cosh\mu)^{8/\triangle}\ve +\vartheta (\ve^2)  \\
\bar{A}_{\scriptscriptstyle{II}} e^{-\bar{m}\rho} u_-(\rho)
+\bar{B}_{\scriptscriptstyle{II}}e^{\bar{m}\rho}u_+(\rho) ,
                    \end{array}
             \right.    \\
\label{solIIIvp}
\vp_{\scriptscriptstyle{III}} & \sim & \left\{\begin{array}{l}
\ve\ln\ve -\ve +\vartheta (\ve^2\ln\ve) \\
\bar{A}_{\scriptscriptstyle{III}} e^{-\bar{m}\rho} u_-(\rho)
+\bar{B}_{\scriptscriptstyle{III}}e^{\bar{m}\rho}u_+(\rho) ,
                    \end{array}
             \right.  \,\,\,\,
\vp_{\scriptscriptstyle{IV}} \sim \left\{\begin{array}{l}
-\ve^{-1}-\fr{\bar{m}^2}{\td}(\cosh\mu)^{8/\triangle}\ln\ve
+\vartheta (\ve^2)\\
\bar{A}_{\scriptscriptstyle{IV}} e^{-\bar{m}\rho} u_-(\rho)
+\bar{B}_{\scriptscriptstyle{IV}}e^{\bar{m}\rho}u_+(\rho) ,
                    \end{array}
             \right.
\eeqa
where upper lines are for $\rho \rightarrow 1$ and bottom lines for
$\rho \rightarrow \infty$. All solutions are simply linear
superpositions of these modes. Note that, for the mode $IV$,
$\vp_{\scriptscriptstyle{IV}} \not= \psi_{\scriptscriptstyle{IV}}$ as one
approaches to the horizon whereas for other three modes $\vp = \psi$.

Now let us turn to the question of boundary conditions for
solutions $\psi$ and $\vp$. Since they are linearized
perturbations, they should be ``small'' at any positions outside
the event horizon. Thus, at the spatial infinity we require both
$\psi$ and $\vp$ should be the decaying one, that is,
exponentially decreasing. In the vicinity of the horizon, the
metric itself is regular in Kruskal coordinates if, and only if,
the perturbation is bounded and $\vp = \psi$ at the
horizon~\cite{Reall:2001ag}. All modes except for the mode $IV$
above satisfy these conditions. However, it is very important to
notice that this is not enough for the regularity of linearized
perturbations. Note that the second derivative of the mode $III$
is proportional to $\ln \ve$. Accordingly some curvature quantity,
for instance, the perturbation of the Ricci scalar curvature
associated with such mode, becomes singular at the horizon. In
other words, the mode solution $III$ produces curvature
singularity even if the metric perturbation itself is regular at
the horizon. Mode solutions $I$ and $II$ do not produce any
curvature singularity at the horizon. By requiring regular
curvature at the horizon in addition, therefore, the boundary
conditions we impose for regular perturbations are that both
$\psi$ and $\vp$ are linear combinations of modes $I$ and $II$
only near the horizon and that they should decay at the spatial
infinity.

For various values of parameters in Eqs.~(\ref{varphif},\ref{psif}),
we have searched for regular solutions satisfying boundary conditions
described above by using Mathematica. In more detail,
we start from a regular solution at the horizon in the following form
\beq
\psi = C\psi_{\scriptscriptstyle{I}} +E\psi_{\scriptscriptstyle{II}},
\qquad \vp = C\vp_{\scriptscriptstyle{I}} +E\vp_{\scriptscriptstyle{II}}.
\label{regsol}
\eeq
At the spatial infinity, they will become
\beqa
\label{asympsolpsi}
\psi & \sim & \left( C A_{\scriptscriptstyle{I}} +E
A_{\scriptscriptstyle{II}} \right) e^{-\bar{m}\rho} v_-(\rho)
+\left( C B_{\scriptscriptstyle{I}} +E B_{\scriptscriptstyle{II}}\right)
e^{\bar{m}\rho} v_+(\rho),  \\
\vp & \sim & \left( C \bar{A}_{\scriptscriptstyle{I}} +E
\bar{A}_{\scriptscriptstyle{II}} \right) e^{-\bar{m}\rho}
u_-(\rho) +\left( C \bar{B}_{\scriptscriptstyle{I}}
+E\bar{B}_{\scriptscriptstyle{II}} \right)e^{\bar{m}\rho}
u_+(\rho) .
\label{asympsolvphi}
\eeqa
By solving
Eqs.~({\ref{varphif},\ref{psif}}) numerically, one can check if
there exists any combination of constants, $C$, $E$, and $\bar{m}$
such that coefficients of the exponentially growing parts for both
$\psi$ and $\vp$ at the spatial infinity vanish in
Eqs.~(\ref{asympsolpsi},\ref{asympsolvphi}), {\it i.e.},
\beq
C
B_{\scriptscriptstyle{I}} +E B_{\scriptscriptstyle{II}} =0, \qquad
C \bar{B}_{\scriptscriptstyle{I}} +E
\bar{B}_{\scriptscriptstyle{II}} =0.
\label{regcond}
\eeq
However,
we have used a different, but equivalent method that turned out to
be more stable and efficient in actual numerical search. For
instance, one just needs to check only for varying $\bar{m}$ as
shall be shown below. Defining
\beq
P(\bar{m},\mu,a,\td,D) =
B_{\scriptscriptstyle{I}} \bar{B}_{\scriptscriptstyle{II}}
-B_{\scriptscriptstyle{II}} \bar{B}_{\scriptscriptstyle{I}} ,
\eeq
the existence of non-trivial constants $C$ and $E$ satisfying Eq.
(\ref{regcond}) is guaranteed only if $P=0$. Thus, having given
the initial data as in Eqs.~(\ref{solI},\ref{solIvp}) near the
horizon, we solve the coupled equations
Eqs.~({\ref{varphif},\ref{psif}}) numerically, and evaluate the
quantity $P(\bar{m},\mu,a,\td,D)$ by using numerical values for
$\psi_{\scriptscriptstyle I,II}$ and $\vp_{\scriptscriptstyle
I,II}$ at sufficiently large $\rho$. For given other parameters of
black $p$-branes, we vary the Kaluza-Klein mass $\bar{m}$ only and
search for the value $\bar{m}^*$ at which $P=0$. This can be
achieved by finding out a $\bar{m}^*$ around which the function
$P$ above changes its sign. If there exists such $\bar{m}^*$, this
mode is indeed the threshold unstable mode.\footnote{In the actual
numerical calculations, we used the sign change of
$\psi_{\scriptscriptstyle{II}}/\psi_{\scriptscriptstyle{I}}
-\vp_{\scriptscriptstyle{II}}/\vp_{\scriptscriptstyle{I}}$ for
$\rho \gg 1$, which becomes equivalent when
$B_{\scriptscriptstyle{I},\scriptscriptstyle{II}},\,
\bar{B}_{\scriptscriptstyle{I},\scriptscriptstyle{II}} \not= 0$.}

\begin{figure}[tbp]
 \centerline{\epsfysize=60mm\epsfxsize=80mm\epsffile{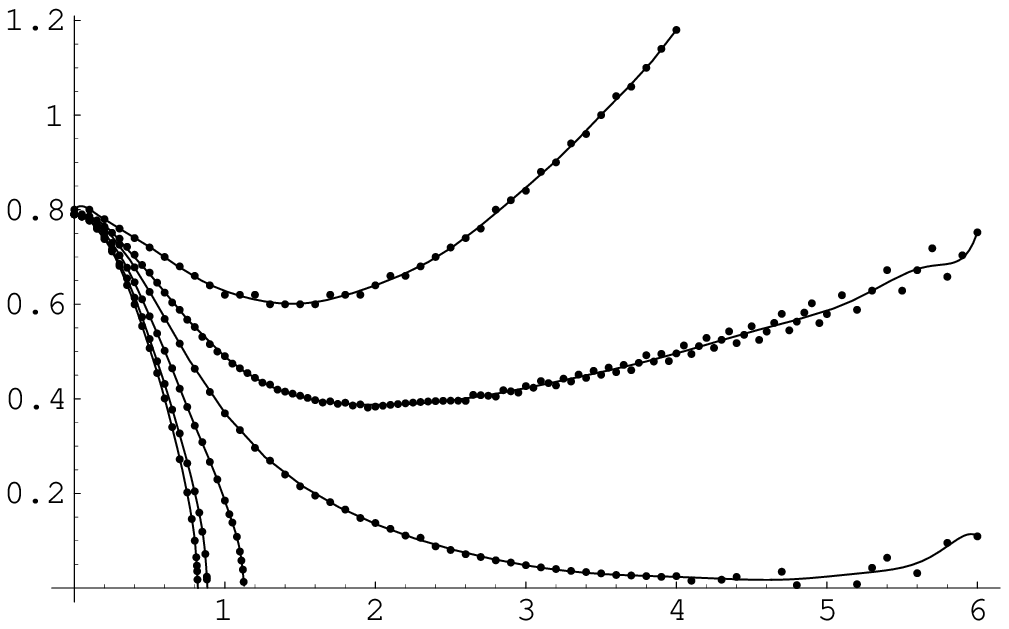}
  \epsfysize=60mm\epsfxsize=80mm\epsffile{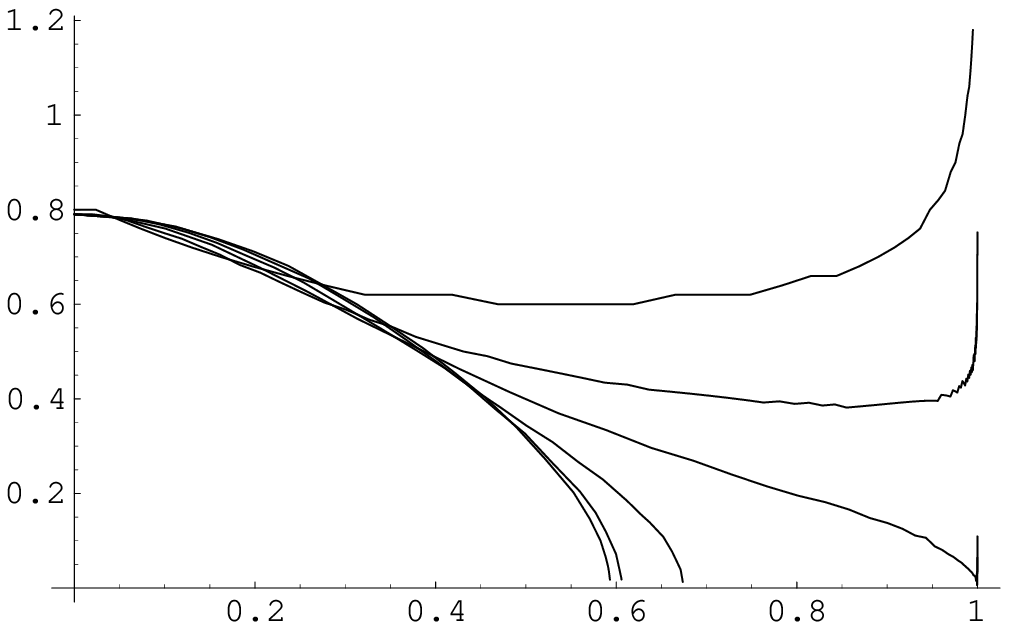}
   \put(-330,193){${\bf p=4}$}
   \put(-450,180){$m^*$}
   \put(-313,140){$\scriptstyle a=3$}
   \put(-317,63){$\scriptstyle a=2$}
   \put(-341,23){$\scriptstyle a=3/2\, ({\rm GL})$}
   \put(-404,32){$\scriptstyle a=1$}
   \put(-411,21){$\scriptstyle a=1/2\, ({\rm D}4)$}
   \put(-432,15){$\scriptstyle a=0$}
   \put(-448,5){$\scriptstyle 0$}
   \put(-233,10){$\mu$}
   \put(-216,5){$\scriptstyle 0$}
   \put(-2,10){$q$}
   \put(-220,180){$m^*$}
   \put(-120,94){$\scriptstyle a=3$}
   \put(-107,75){$\scriptstyle a=2$}
   \put(-95,58){$\scriptstyle a=3/2\, ({\rm GL})$}
   \put(-87,40){$\scriptstyle a=1$}
   \put(-87,24){$\scriptstyle a=1/2\, ({\rm D}4)$}
   \put(-106,14){$\scriptstyle a=0$}
    }
\vspace{0.5cm}
\caption{Behavior of threshold masses for black
$4$-branes in $D=10$ at various values of $a$ with fixed mass
density $M=2^5$. $m^* \simeq 1.581/r_{\scriptscriptstyle{H}}\simeq
0.791$ with $r_{\scriptscriptstyle{H}}=2$ at $\mu=0$. $\mu_{\rm
cr} \simeq 0.818\,(0.8184)$, $0.881\,(0.8814)$, and
$1.125\,(1.1254)$ for cases of $a=0$, $1/2$, and $1$,
respectively. Here critical values obtained from the GM conjecture
are denoted in parentheses. On the righthand side the same data
are plotted in terms of the non-extremality parameter $q$. $\mu=6$
corresponds to $q \simeq 0.99996$, $0.99995$, $0.99991$ for
$a=3/2$, $2$, $3$.}
\label{fig1}
\end{figure}

Once such $\bar{m}^*=\bar{m}^*(\mu,a,\td,D)$ is obtained
numerically, the threshold mass
$m^*=\bar{m}^*/r_{\scriptscriptstyle{H}}=\bar{m}^*/k^{1/\td}$ can
be evaluated by fixing $r_{\scriptscriptstyle{H}}$ or $k$. Recall
that a black $p$-brane in $D$ dimensions is characterized by two
physical quantities $Q$ and $M$, or equivalently $r_H$ and $\mu$,
and one coupling parameter $a$. We would like to see how the
stability behaves as the charge $Q$ of a black brane increases
when its mass density $M$ is kept same.\footnote{In the Gregory
and Laflamme's work, they fix the size of the outer horizon $r_ 0$
in string frame and vary the inner horizon size $r_i$. Here $r_0^{
\td}= r_ {\scriptscriptstyle{H}}^{\td} \cosh^2\mu$ and $r_i^{
\td}= r_ {\scriptscriptstyle{H}}^{\td} \sinh^2\mu$.} We have
chosen $M$ such that threshold masses at $\mu=0$ for the GL cases
reproduce those in Gregory and Laflamme's work
\cite{Gregory:1994bj,Gregory:1994tw}, {\it i.e.},
$r_0=r_{\scriptscriptstyle{H}}=2$. A non-extremality parameter can
be defined as
\beq
q \equiv
\fr{Q}{Q_{\rm{\scriptscriptstyle{max}}}}
  = \fr{2}{\sr{\triangle}}\fr{Q}{M}
  = \fr{\sinh 2\mu}{(\td +1)\triangle/2\td +2\sinh^2\mu}.
\label{qvalue}
\eeq
Here $Q_{\rm max}$ is the maximum charge
density allowed for a given mass density $M$. Since $(\td
+1)\triangle/2\td \geq 1$, $q$ is a monotonic function of $\mu$
and $q=1$ is the extremal case. Notice that as $\mu$ or $q$
increases, the horizon radius $r_{\scriptscriptstyle{H}}=
k^{1/\td}$ decreases according to Eq. (\ref{MC}) and becomes zero
for the extremal brane when $M$ is fixed. For black branes having
different mass density $M'$, one can get
\beq
m^*(\mu;M')=\left(\fr{M}{M'}\right)^{1/\td} m^*(\mu ;M) .
\eeq

Behaviors of some of threshold masses we have obtained are
illustrated in Fig.~\ref{fig1}. There we have considered black
$4$-branes in $D=10$ and the mass density $M=2^5$. Our numerical
results show that threshold masses at $\mu =0$ are nonvanishing
for any values of $a$, implying that all uncharged black
$4$-branes in Eq.~(\ref{conmetric}) are unstable under linearized
perturbations associated with Kaluza-Klein massive modes with $0 <
m < m^*$ as expected. In other words, instability occurs for small
perturbations whose wavelengths in spatial worldvolume directions
are in the range of $\lambda_* (= 2\pi/m^*) < \lambda < \infty$.
The value of threshold masses at $\mu=0$ can be seen to be
independent of $a$ as it should be so since the $a$-dependence of
Eqs.~(\ref{varphif},\ref{psif}) disappears as $\mu \rightarrow 0$.
Numerically we find $m^* \simeq 0.791$ at $\mu=0$, which agrees
with the numerical result obtained by Gregory and Laflamme
\cite{Gregory:1994bj}.

When black branes get charged, the background dilaton field
becomes nontrivial ({\it e.g.}, $\phi \not= 0$) and could play
some important role. As can be seen in Fig.\,\ref{fig1}, the
stability behaves very differently depending on the coupling
parameter $a$ between dilaton and gauge fields in Einstein frame
Eq. (\ref{action1}). We summarize its behavior into three types,
{\it i.e.}, $a < 3/2$, $a = 3/2$ and $a > 3/2$.

\begin{figure}[tbp]
 \centerline{\epsfysize=60mm\epsfxsize=80mm\epsffile{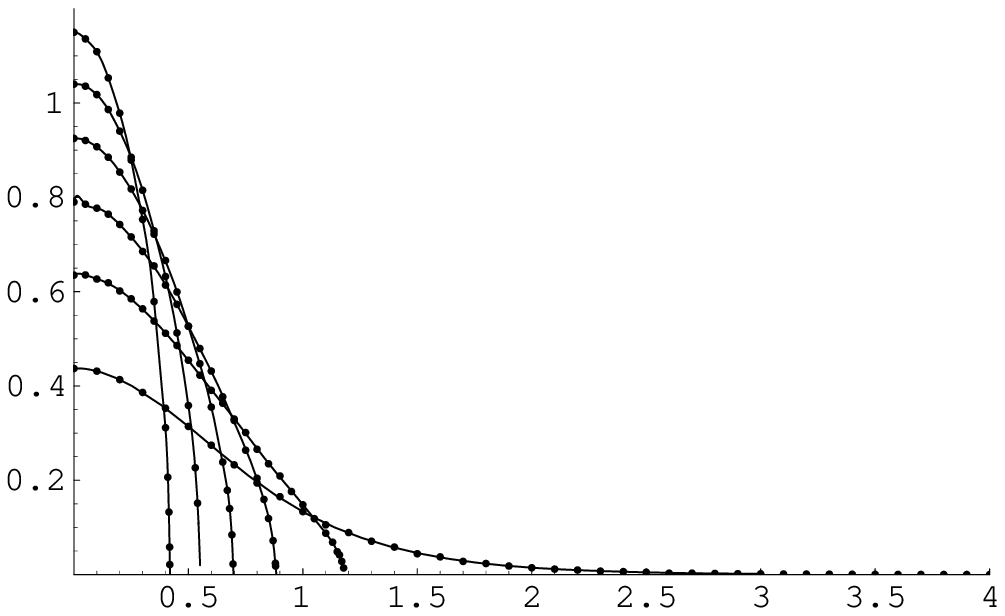}
  \epsfysize=60mm\epsfxsize=80mm\epsffile{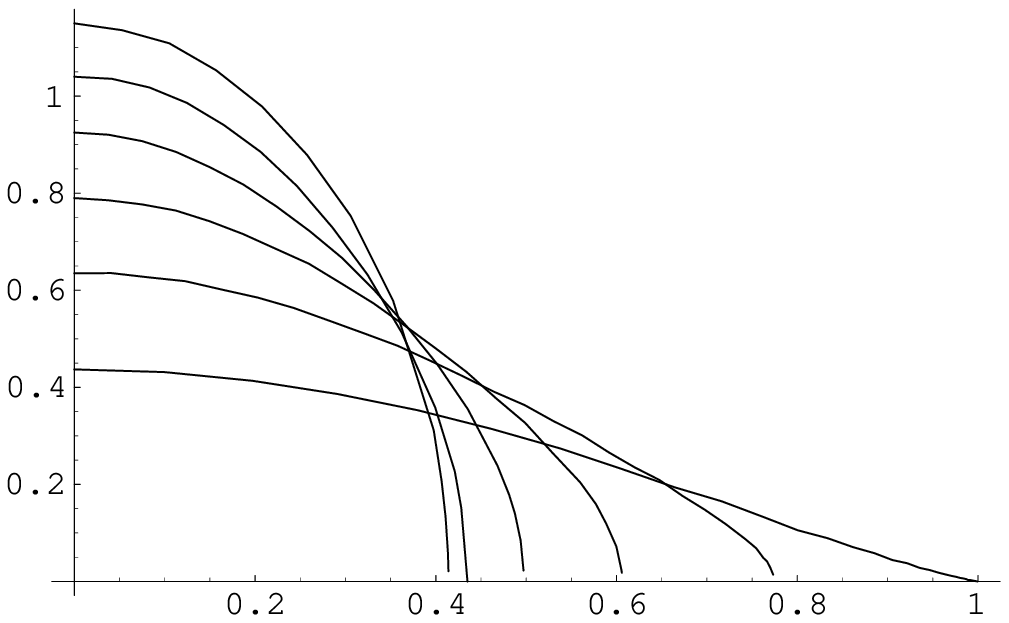}
   \put(-335,193){${\bf |a|=1/2}$}
   \put(-450,180){$m^*$}
   \put(-449,4){$\scriptstyle 0$}
   \put(-233,12){$\mu$}
   \put(-430,153){$\scriptstyle p=1$}
   \put(-422,120){$\scriptstyle p=2\, ({\rm D}2)$}
   \put(-415,97){$\scriptstyle p=3$}
   \put(-407,72){$\scriptstyle p=4\, ({\rm D}4)$}
   \put(-391,43){$\scriptstyle p=5$}
   \put(-362,24){$\scriptstyle p=6\, ({\rm GL})$}
   \put(-217,4){$\scriptstyle 0$}
   \put(-2,12){$q$}
   \put(-220,180){$m^*$}
   \put(-200,168){$\scriptstyle p=1$}
   \put(-200,152){$\scriptstyle p=2\, ({\rm D}2)$}
   \put(-200,137){$\scriptstyle p=3$}
   \put(-200,119){$\scriptstyle p=4\, ({\rm D}4)$}
   \put(-200,100){$\scriptstyle p=5$}
   \put(-200,73){$\scriptstyle p=6\, ({\rm GL})$}
    }
\vspace{0.5cm} \caption{Behavior of threshold masses for black
$p$-branes with fixed $M$ in $D=10$ in the theory of $|a|=1/2$. At
$\mu=0$, $r_{\scriptscriptstyle{H}}=2$ and
$m^*=\bar{m}^*/r_{\scriptscriptstyle{H}} \simeq 1.153$, $1.044$,
$0.925$, $0.791$, $0.635$, $0.440$ for $p=1,$ $\cdots$, $6$.
Critical values for transition points are $q_{\rm cr} \simeq
0.413$, $0.435$, $0.497$, $0.606$, $0.773$, $>0.9993$ for $p=1,
\cdots, 6$. } \label{fig2}
\end{figure}

In the work by Gregory and Laflamme
\cite{Gregory:1994bj,Gregory:1994tw}, it has been shown that the
black brane instability decreases monotonically as a black brane
approaches to its extremal point.\footnote{By decreasing
instability, we mean that the threshold mass is decreasing. Thus,
the instability actually shrinks in parameter range of $\Omega$
and $m$. Consequently, smaller values of $\Omega$ imply that
instability modes grow less rapidly in time. So, in this sense we
may say a black brane having larger charge is less unstable than
those having smaller charges in the case considered by Gregory and
Laflamme.} It is also noticed that numerical instability starts to
occur near the extremal point, but the extremal brane was shown to
be stable by analyzing its stability separately
\cite{Gregory:1994tw}. The case for which numerical study was
performed explicitly by Gregory and Laflamme corresponds to the GL
case in Fig.\,\ref{fig2}, {\it i.e.}, $a=-1/2$ and $p=6$. Our
result confirms theirs up to $\mu \simeq 4$ or $q \simeq 0.9993$.
When $\mu \,\, {\scriptstyle \gtrsim} \,\, 4$, numerical
instability starts to occur as already observed in Refs.
\cite{Gregory:1994bj,Gregory:1994tw}. For the GL case in
Fig.\,\ref{fig1} ({\it i.e.}, $a=3/2$ and $p=4$), the stability
behavior is similar at least up to $\mu \simeq 4$ or $q \simeq
0.99799$. Our data for $\mu \,\, {\scriptstyle \gtrsim} \,\, 4$
are not reliable due to numerical instability.

For cases of $a=0$, $1/2$, and $1$, we find that threshold masses
become zero at certain finite values of $\mu_{\rm cr} \simeq
0.818$, $0.881$, and $1.125$, respectively. Thus the
Gregory-Laflamme instability disappears as $\mu \rightarrow
\mu_{\rm cr}$ for such cases and so corresponding black $4$-branes
become in fact {\it stable}, at least under spherically symmetric
linearized perturbations, for $\mu_{\rm cr} < \mu \leq 6$,
presumably all the way down to the extremal point ({\it i.e.},
$\mu \sim \infty$). The same data are plotted in terms of the
non-extremality parameter $q$ on the righthand side. Corresponding
critical values for transitions in the stability are $q_{\rm cr}
\simeq 0.593$, $0.606$, $0.674$ for $a=0$, $1/2$, $1$,
respectively. This critical value $q_{\rm cr}$ seems to increase
monotonically from a minimum one for $a=0$ to unit for the GL
case. Therefore one sees that there indeed exist some charged
black branes which are stable even far from the extremality.

For cases of $a=2$ and $3$, threshold masses decrease as black
4-branes get charged as before, but interestingly they start to
increase again at around $\mu \simeq 2$ ({\it i.e.}, $q \simeq
0.9972$) and $\mu \simeq 1.5$ ({\it i.e.}, $q \simeq 0.9950$),
respectively. This turning point becomes smaller as $a$ increases.
Thus it shows that the instability does {\it not} always reduce
down as a black brane gains more charge. Such stability behavior
in the presence of charge never has been expected in the
literature as far as we know. Moreover, the threshold mass seems
to diverge as the black brane approaches to the extremal point as
can be seen in Fig.\,\ref{fig1} up to $q \simeq 0.99995$,
$0.99991$ for $a=2$, $3$. As shown below, however, extremal black
branes are stable in these cases as well.

Now let us check the stability of extremal black branes. These
cases are expected to be stable since they correspond to BPS
ground states in string theory. Our numerical analysis cannot be
directly used to show its stability since numerical instability
occurs as the extremality is being approached. Instead, we treat
the extremal case separately. As one can show explicitly,
equations for linear perturbations in the extremal case are
obtained simply by putting $k=0$ and $\mu=\infty$ with
$ke^{2\mu}=4c$ fixed in Eqs.\,(\ref{varphif},\ref{psif}),
\beqa
\label{vphiex}
\mbox{} & & \vp'' +\fr{\td+1}{r}\vp' -m^2\left(
1+\fr{c}{r^{\td}}
\right)^{4/\triangle} \vp =0 ,   \\
\label{psiex}
& & (r^{\td}+c) \psi'' +\fr{1}{r}\Bigg[ (\td +3)r^{\td}
-4c \Big(\fr{\td}{\triangle} +\fr{\td -3}{4}\Big)\Bigg] \psi'
-\Bigg[ m^2(r^{\td} +c)(1+\fr{c}{r^{\td}})^{4/\triangle} \nonumber \\
& & \qquad\qquad\quad +\fr{2c\td}{\triangle r^2}\Big( 2\td^2 +(\td
+2)(a^2-\fr{2\td^2}{D-2}) \Big)\Bigg] \psi -\fr{2c\td}{\triangle
r^2}\Bigg( a^2-\fr{2\td^2}{D-2} \Bigg) \vp =0 .
\eeqa
The boundary
conditions are simply the regularity of $\vp$ and $\psi$, that are
different from those of nonextremal cases, and the asymptotic
solutions near the event horizon are also different and cannot be
obtained simply by taking the extremal limit of those in
Eqs.\,(\ref{solI}-\ref{solIIIvp}). This difference comes because
we cannot ignore the terms including $k$ at the near horizon as
long as $k$ is non-vanishing. Now it is straightforward to see
that the only regular solution of Eq.\,(\ref{vphiex}) is $\vp =0$
since the coefficient of $\vp$ is negative definite. Similarly,
the only regular solution for $\psi$ is zero. Therefore all
extremal black branes considered in this paper are stable at least
under $s$-wave fluctuations, although the threshold mass seems to
diverge for $a>3/2$ as the extremality is being approached.

For black $p$-branes with different $p$ ($\neq 4$), the basic
stability behavior is essentially the same as the above. Only the
values of threshold masses and critical $\mu$ change slightly.
Fig.\,\ref{fig2} shows our results for various $p$-branes in the
theory of $|a|=1/2$. The mass density is chosen to be
$M=(8-p)2^{7-p}$, giving $r_{\scriptscriptstyle{H}}=2$ at $\mu=0$.
If results for other values of $a$ are added, one can see similar
patterns as in Fig.\,\ref{fig1} for each $p$-brane. Black
$p$-branes with $p=1, \cdots, 5$ becomes stable if $\mu \geq
\mu_{\rm cr}$ whereas the instability of the $6$-brane persists
all the way down to at least $\mu=4$ ({\it i.e.}, $q=0.9993$).

\begin{figure}[tbp]
 \centerline{\epsfxsize=85mm\epsffile{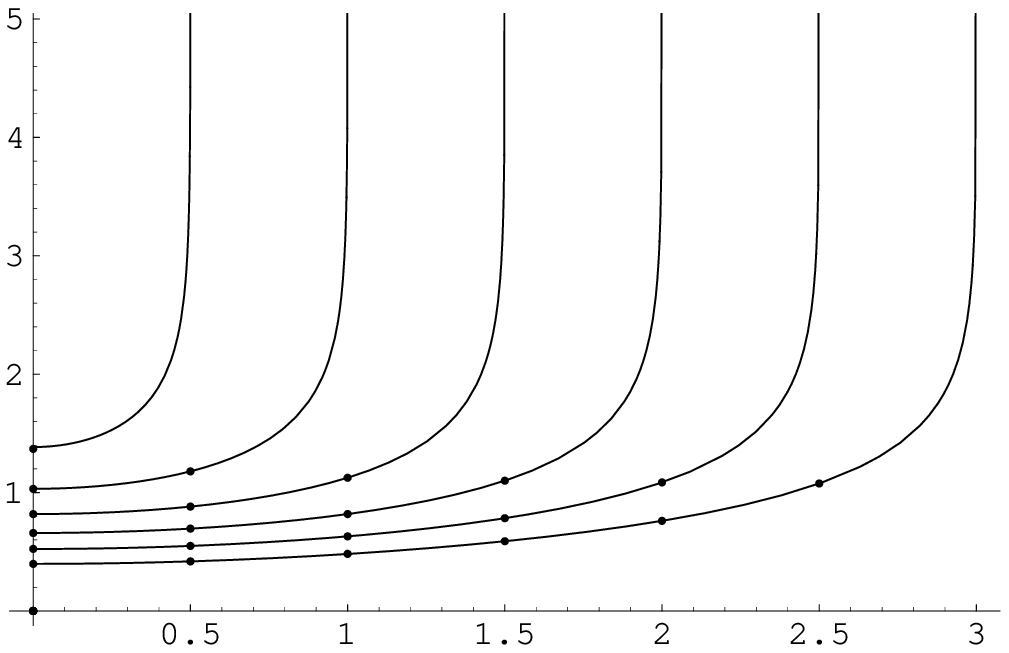}
   \put(-260,150){$\mu_{\rm cr}$}
   \put(5,5){$|a|$}
   \put(-193,125){$\scriptstyle p=6$}
   \put(-155,125){$\scriptstyle p=5$}
   \put(-115,125){$\scriptstyle p=4$}
   \put(-78,125){$\scriptstyle p=3$}
   \put(-40,125){$\scriptstyle p=2$}
   \put(0,125){$\scriptstyle p=1$}
   \put(-156,22){$\scriptstyle {\rm F}1,{\rm D}1$}
   \put(-195,24){$\scriptstyle {\rm D}2$}
   \put(-195,34){$\scriptstyle {\rm D}4$}
   \put(-157,138){$\scriptstyle {\rm NS}5$}
   \put(-157,146){$\scriptstyle {\rm D}5$}
   \put(-158,156){$\scriptstyle \uparrow$}
   \put(-160,164){$\scriptstyle \infty$}
    }
\vspace{0.5cm}
\caption{Critical values of the parameter $\mu$ for
various black $p$-branes in $D=10$ at which the threshold mass
vanishes $m^*=0$. The solid lines are obtained from the
Gubser-Mitra conjecture and the black dots from our numerical
results for several values of $a$.}
\label{fig3}
\end{figure}

Now it will be very interesting to see how well the classical
stability behavior of black branes described above agrees with
that predicted by the local thermodynamic behavior through the GM
conjecture. First of all, let us consider critical values
$\mu_{\rm cr}$ beyond that black branes become stable classically
from being unstable.
\beq
\begin{array}{l|llllll}
\qquad p   & \quad 1 & \quad 2 & \quad 3 & \quad 4 & \quad 5 & \,\, 6 \\
 \hline
 \mu_{\rm cr}\,({\rm Num.}) & 0.418 & 0.549 & 0.695 & 0.881 & 1.178 & >4 \\
 \mu_{\rm cr}\,({\rm GM})   & 0.4186 & 0.5493 & 0.6954 & 0.8814 & 1.1791 &
 \infty \\
\end{array}
\eeq
The table above shows the results for the case of $|a|=1/2$
in Fig.\,\ref{fig2}. One can see that they are in very good
agreement. Other cases we have checked are marked with black dots
in Fig.\,\ref{fig3}. Solid lines are obtained by using
Eq.\,(\ref{criticalmu}) in the GM conjecture. All critical values
of $\mu$ obtained numerically in our classical perturbation
analysis agreed well with those in the GM conjecture. Based on
such good agreement, this diagram shows how the classical
stability under small perturbations will behave in general. That
is, for a black brane with given dimension $p$ for the spatial
worldvolume, this brane will be unstable ({\it i.e.}, $\mu_{\rm
cr} =\infty$) under small perturbations if $|a| \geq a_{\rm
cr}=(D-3-p)/\sr{(D-2)/2}$. If $|a| < a_{\rm cr}$, the brane is
still unstable for $\mu < \mu_{\rm cr}(a,p,D)$, but becomes stable
for $\mu \geq \mu_{\rm cr}$. As can be seen in Fig.\,\ref{fig3}
for $D=10$, critical values $\mu_{\rm cr}$ increase monotonically
as $|a|$ or $p$ increases. Critical values for various black brane
solutions of the type II supergravity are marked in
Fig.\,\ref{fig3} explicitly.

\section{Discussion}

To conclude, we have investigated the stability of magnetically
charged black brane solutions for low energy string theory in
Eq.~(\ref{action2}) under small perturbations. It turns out that
all uncharged black branes in our consideration are unstable under
linearized perturbations. When the black brane gets charged,
however, the stability behavior depends on how strongly the
$n$-form field couples to the dilaton field. In more detail, our
results seem to show that, when the coupling is weak enough in the
theory in Eq.\,(\ref{action1}) ({\it e.g.}, $|a| < a_{\rm cr}$),
black branes become stable as they get charged enough ({\it e.g.},
$\mu \geq \mu_{\rm cr}$), even before they reach to the extremal
point. When the coupling is strong enough ({\it e.g.}, $|a| \geq
a_{\rm cr}$), however, black brane solutions of this theory are
always unstable. Moreover the instability starts to increase again
as the charge is larger than a certain value and seems to diverge
near the extremality. For example, F1, D1, D2, D4 black branes of
the type II supergravity could be stable classically for large
charge, whereas NS5, D5 black brane solutions are always unstable
all the way down to the extremal point. The case of $a=a_{\rm cr}$
is the boundary between these two categories, which is the case
Gregory and Laflamme studied, and so the instability monotonically
reduces down to zero at the extremal point. All extremal black
branes are shown to be stable separately. It has also been shown
that our results for the classical stability agree well with the
qualitative behavior predicted by the local thermodynamic
stability through the Gubser-Mitra conjecture. That is, the
critical values for $a_{\rm cr}$ and $\mu_{\rm cr}$ obtained
numerically in our classical perturbation analysis agree very well
with those in Eqs.\,(\ref{criticala},\ref{criticalmu}) for the
sign change of the specific heat for the black brane being
regarded as a thermal body.

In the classical perturbation analysis above, we have considered
only spherically symmetric perturbations. Thus, even if it turned
out that there exists no instability mode for a black brane
solution with $\mu > \mu_{\rm cr}$ in the analysis above, this
does not necessarily mean such black brane is stable under small
perturbations. There might exist some instability mode when we
considered all non-$s$-wave perturbations as well. However, we
give some evidence that the only possible instability mode comes
from $s$-wave fluctuations as follows: Although it was in a
different context, one can find in Ref.\,\cite{GPY} that the
higher angular momentum fluctuations for Schwarzschild black brane
backgrounds do not produce unstable modes. As $\mu \rightarrow 0$,
since the black brane solutions considered in the present paper
become Schwarzschild black branes and the metric perturbation
equations become completely decoupled from others, one can easily
see that the $s$-mode instability is the only instability for {\it
uncharged} black branes in this paper. Now one can apply the same
argument as in Ref. \cite{Gregory:1994tw}. When charge is added,
we observed that there is a stabilizing influence for the $s$-wave
perturbations if $|a| \leq a_{\rm cr}$. Therefore, it is not
expected that higher angular momentum modes exhibit instability in
{\it charged} black branes since they do not give it even for the
uncharged case.

The spatial worldvolume is assumed to be non-compact in the
description above. Consequently, the KK mass spectrum is
continuous since there are translational symmetries in spatial
worldvolume directions. What will happen on the stability behavior
if the worldvolume is compactified with a scale $L$? In this case,
the KK mass becomes discrete and has a minimum mass, {\it e.g.},
$m_{\rm min} =2\pi /L$. Thus, KK masses smaller than the minimum
({\it i.e.}, $0 \leq m < m_{\rm min}$) are not allowed due to the
compactification. Accordingly, even if threshold masses obtained
by solving perturbation equations for black branes are
nonvanishing, such black branes are actually stable provided that
$m^* < m_{\rm min}$, or equivalently that the compactification
scale is small enough ({\it i.e.}, $L < 2\pi /m^*$)
\cite{Gregory:vy,Gregory:1994bj,Hubeny:2002xn}.

Although critical values for transition points are in good
agreement between the classical stability analysis and the local
thermodynamic stability through the GM conjecture, there are some
other aspects that might be in disagreement. For instance, one can
see in cases of $|a| \leq a_{\rm cr}$ that the magnitude of the
specific heat increases whereas the black brane instability
reduces as the charge increases with fixed mass density. Moreover,
the specific heat is divergent at the transition point ({\it
i.e.}, $\mu =\mu_{\rm cr}$). In the classical perturbation
analysis, however, there does not seem to exist any singular
behavior around $\mu =\mu_{\rm cr}$. For given $\mu$ in
Fig.\,\ref{fig1}, we found the threshold mass increases as $a$
increases. One might wonder if the magnitude of the specific heat
also increases. One can see that it actually decreases
monotonically, even for a fixed $Q/M$ instead of a fixed $\mu$.
Maybe these discrepancies for some details in stability behavior
simply indicate that one needs to find out some appropriately
modified thermodynamic quantity, other than the specific heat, in
order to compare other details between classical and thermodynamic
stability analyses for black branes.

Finally, it would be very interesting to understand why the
instability starts to increase again as the extremality is being
approached for black branes in theories of $a > a_{\rm cr}$.
Further work is required.

\section*{Acknowledgments}

GK would like to thank H.S. Reall for useful discussions. This
work was supported by JSPS (Japanese Society for Promotion of
Sciences) Postdoctoral Fellowships. Y. Lee was supported in part
by the BK21 Project of the Ministry of Education, Korea.

\newcommand{\J}[4]{#1 {\bf #2} #3 (#4)}
\newcommand{\andJ}[3]{{\bf #1} (#2) #3}
\newcommand{\AP}{Ann.\ Phys.\ (N.Y.)}
\newcommand{\MPL}{Mod.\ Phys.\ Lett.}
\newcommand{\NP}{Nucl.\ Phys.}
\newcommand{\PL}{Phys.\ Lett.}
\newcommand{\PR}{Phys.\ Rev.\ D}
\newcommand{\PRL}{Phys.\ Rev.\ Lett.}
\newcommand{\PTP}{Prog.\ Theor.\ Phys.}
\newcommand{\hep}[1]{ hep-th/{#1}}
\newcommand{\hepp}[1]{ hep-ph/{#1}}
\newcommand{\hepg}[1]{ gr-qc/{#1}}


\begin{thebibliography}{999}
\bibitem{footnote2}
For massless modes, the perturbation equations become equivalent
to those for black hole bacgrounds which are expected to be
stable.
\end{thebibliography}
\end{document}